\documentclass[prd,aps,twocolumn]{revtex4}
\usepackage{url}
\usepackage{natbib}
\usepackage{color}
\usepackage{amsmath}
\usepackage{graphicx} 
\begin{document}
\title{Pulsar timing array constraints on the cosmological magnetic field from quark confinement epoch}
\author{Theo Boyer$^1$, Andrii Neronov$^{1,2}$ }
\affiliation{$^{1}$Universit\'e Paris Cit\'e, CNRS, Astroparticule et Cosmologie, 
F-75013 Paris, France}

\affiliation{$^{2}$Laboratory of Astrophysics, \'Ecole Polytechnique F\'ed\'erale de Lausanne, CH-1015 Lausanne, Switzerland}
\begin{abstract}
Recent evidence for the stochastic gravitational wave backgorund reported by the pulsar timing arrays (PTA) can be interpreted as a signal from the cosmological phase transition. We use up-to-date models of the gravitational wave power spectra to compare constraints on the parameters of the phase transition for the three different available PTA measurements and to work out a refined estimate of the cosmological magnetic field that should result from this transition. We find that the PTA data, combined with a constraint from the abundance of primordial black holes, are consistent with a possibility of a moderate strength first-order phase transition during quark confinement and yield a rather precise prediction for the initial parameters of the magnetic field, with the magnetic field energy density in near equipartition with photon energy density and correlation length close to one co-moving parsec.
\end{abstract}

\maketitle

\section{Introduction}

Pulsar Tiniming Array (PTA) gravitational wave detectors,  NANOGrav \cite{NANOGrav:2023gor}, EPTA \cite{EPTA:2023fyk}, PPTA \cite{Reardon:2023gzh}, and CPTA \cite{Xu:2023wog}, have recently revealed evidence for a nano-Hertz Stochastic Gravitational Wave Background (SGWB). Supermassive black hole binaries in the centers of galaxies may naturally generate such SGWB \cite{Burke-Spolaor:2018bvk}, but the detected signal appears to be in mild tension with the model predictions for the binary supermassive black hole population  \cite{Sato-Polito:2023gym,NANOGrav:2023hvm}. An alternative source of the nHz SGWB is a cosmological first-order phase transition at the Quantum Chromodynamics (QCD) energy scale \cite{PhysRevD.30.272,Kosowsky:1992vn,Caprini:2007xq,NANOGrav:2021flc}. Such phase transition proceeds through nucleation, growth and collisions of bubbles of new phase that generate stress-energy tensor sourcing the gravitational waves. Plasma motions induced by the bubble dynamics also create magnetic fields \cite{Sigl:1996dm} that can also be a source of gravitational waves through the Magneto-Hydro-Dynamical (MHD) turbulence \cite{Neronov:2020qrl,RoperPol:2022iel,Brandenburg:2021tmp,EPTA:2023xxk}. 
Within such interpretation, measurement of the parameters of the SGWB can be used to estimate the strength and correlation length of magnetic field present in the Universe during the QCD epoch \cite{Neronov:2020qrl}. Previous estimates of the field magnetic parameters based on the earlier NANOGrav data  suggested that such field would  evolve into a magnetic field that would affect the dynamics of recombination strongly enough to change the estimate of the Hubble parameter from the Cosmic Microwave Background data and alleviate the Hubble tension problem \cite{Jedamzik:2020krr,Galli:2021mxk,RoperPol:2022iel}. 

The simplest phenomenological models of cosmological phase transitions describe their dynamics in terms of four parameters, the bubble nucleation temperature $T_*$,  the latent heat $\alpha$, the rate $\beta$,  and the velocity of the bubble walls $v_w$. Constraints on  $T_*, \beta$ from the new release of the NANOGrav data have been derived in  Refs.
\cite{NANOGrav:2023hvm,Ellis:2023oxs}, confirming previous estimates that the phase transition temperature is close to the QCD energy scale $T\sim 10^2$~MeV and the time scale of the transition $\tau=\beta^{-1}$ is close to the Hubble scale $t_H=H(T_*)^{-1}$ ($H$ is the Hubble parameter). Implications for the cosmological magnetic field have been considered by the EPTA collaboration \cite{EPTA:2023xxk}, for the possibility of the MHD turbulence driving the production of the gravitational waves. Ref. \cite{RoperPol:2023bqa} has considered an alternative scenario in which the MHD turbulence is instead excited by the plasma motions during the first-order phase transitions. In this case the gravitational wave spectrum has contributions from both the MHD turbulence (like in Ref. \cite{EPTA:2023xxk}) and from the sound waves excited by the bubble collisions (like in \cite{NANOGrav:2023hvm,Ellis:2023oxs}). 

In what follows we apply the formalism of Ref. \cite{RoperPol:2023bqa} to a combination of the new PTA data. We derive a refined estimate of the range of the primordial magnetic field parameters at the QCD epoch that provides and improvement on  previous result  of Ref. \cite{Neronov:2020qrl}. In this previous work, as well as in the analysis of Ref. \cite{EPTA:2023xxk}, the MHD turbulence was considered as the main source to the SGWB. We find that the evidence for the SGWB, interpreted as a signal from the cosmological phase transition implies a rather precise measurement of the correlation length of the cosmological magnetic field. The predictions for the magnetic field strength are less precise because of uncertainty of the fraction of the energy release of the phase transition that goes into MHD turbulence, $\epsilon_{turb}$. However, for a reasonable order-of-magnitude assumption of this fraction, also the initial magnetic field strength is constrained with less than order-of-magnitude uncertainty.

\section{Modelling of the gravitational wave spectra}

Our analysis is based on parameterisations of the SGWB power spectrum components due to the sound waves and MHD turbulence summarised in \cite{RoperPol:2023bqa}. The spectra depend on the five phenomenological parameters introduced above, $\alpha, \beta, T_*, v_w, \epsilon_{turb}$. The SGWB power spectrum has two main contributions: from the sound waves and from the MHD turbulence. Below we provide a summary of the analytical formulae for both components. We make the model calculations available as an online service at the Multi-Messenger Online Data Analysis platform (MMODA, {\url{https://mmoda.io}) and on a generic online analysis platform Galaxy (\url{https://usegalaxy.eu})}.  The service provides the spectra of the two components in a tabular form as a function of the five phenomenological parameters of the phase transition.

\subsection{Sound waves component}

The sound waves gravitational wave spectrum is based on the parameterisations of Ref. \cite{Hindmarsh_2019}. We can express the gravitational waves (GW) spectrum as a broken power law as a function of frequency $f$:

\begin{equation}
    \Omega_{\rm GW}^{\rm sw} (f) = 3\, {\cal B} (\Delta_w) 
    \frac{(K \lambda_* {\cal H}_*)^2}{\sqrt{K} + \lambda_* {\cal H_*}}
    F_{\rm GW}^0 \,
    S_{\rm sw} (s_1, s_2, \Delta_w)
    \label{OmGW_SW}
\end{equation}

where :

\begin{itemize}
    \item ${\cal B} (\Delta_w)$ is a dimensionless parameter, giving the efficiency
    with which shear stress is converted into gravitational
    waves, as described in Ref. \cite{B(delta_w)}. This efficiency depends only on  $\Delta_w = \frac{|v_w-c_s|}{v_w}$, where $c_s=c/\sqrt{3}$ is 
    the speed of sound. The functional form of ${\cal B}$ can be found in \cite{B(delta_w)}.
    
    \item $K$ is the amount of sound wave kinetic energy density as a fraction of the total energy density:
\begin{equation}
    K = \frac{\rho_{\rm K}}{\rho_{\rm vac} + \rho_{\rm rad}^*} =
    \frac{\kappa_v \alpha}{1 + \alpha},
    \label{K_def}
\end{equation}
where $\kappa_v \equiv \rho_{\rm K}/\rho_{\rm vac}$ is the fraction of the vacuum energy released that
is converted to kinetic energy density of the plasma $\rho_{\rm K}$. $\kappa_v$ depends only on $\alpha$ and $v_w$. The full expression for $\kappa_v$ can be found in Ref.  \citep{Espinosa_2010}.
\item $\lambda_*$ is the mean comoving radius of the new phase bubbles at the time of percolation which depends only on $\beta$ and $v_w$ such as :
    \begin{equation}
    \label{lambda}
\lambda_{*}=\frac{(8\pi)^{1/3}\max(v_w,c_s)}{\beta}
    \end{equation}

\item ${\cal H}_*$ is the comoving Hubble rate at the moment of percolation;

    \item $F^0_{GW}$ is the transfer function that takes into account the redshift of the GW energy density such as :
    \begin{equation}
    F_{\rm GW}^0 = \biggl(\frac{a_*}{a_0}\biggr)^4
    \biggl(\frac{H_*}{H_0}\biggr)^2
    \simeq 1.64\, h_0^{-2} \times 
    10^{-5} \,
    \biggl(\frac{100}{g_*}\biggr)^{1\over3}
    \end{equation}
    where $a_*$ is the scale factor at the time of percolation and $g_*$ the number of relativistic degrees of freedom at the time of percolation.

\item $S_{\rm sw}$ is the spectral shape (which is a broken power-law) \cite{SWGW}.
    \begin{equation}
    S_{\rm sw}^{\rm HL} (s_1, s_2, \Delta_w) = \frac{16 \, (1 + \Delta_w^{-3})^{2\over 3} ({\Delta_w} \, s_1)^3}{(1 + s_1^3)^{{2\over 3}} (3 + s_2^2)^{2}}
    \label{spectral_shape_HL}
    \end{equation}
    where $s_1 = \lambda_* f$ and $s_2 = \delta \lambda_* f =\Delta_w \lambda_* f$.
\end{itemize}

\subsection{Turbulence component}

The second component of the total GW spectrum is produced by MHD turbulence. The parameterisation is valid for the situation in which the turbulence if forced starting from the moment of percolation of the bubbles during a certain time interval  $\delta {t_{fin}}$. The numerical simulations of the gravitational wave production by the MHD turbulence \cite{Roper_Pol_2022} suggest that  $\delta {t_{fin}}\simeq 2 \lambda_{*} / u_{*}$,
where $u_*$ is the characteristic velocity scale of the turbulence, either the Alfven velocity, or the velocity of the plasma. It is related to the turbulence energy density, expressed as a fraction of the total energy density, $\Omega_*=\epsilon_{turb}K$ \cite{RoperPol:2023bqa}
\begin{equation}
    u_* = \sqrt{{3\over4}\Omega_*}.
\end{equation}

The turbulent component of the SGWB spectrum is
\begin{equation}
    \Omega_{\rm GW}^{\rm turb} (f) = 3 \, {\cal A} \, \Omega_{*}^2  \, (\lambda_* {\cal H}_*)^2
    F_{\rm GW}^0 \,
    S_{\rm turb} (s_1, s_3).
    \label{eq:OmGWturb}
\end{equation}

where :

\begin{itemize}
    \item ${\cal A}\simeq 1.8\times 10^{-3}$ is a normalization factor \cite{RoperPol:2022iel}
    
\item $S_{turb}$ is the broken power law spectrum \cite{RoperPol:2022iel} 
    \begin{equation}
    S_{\rm turb} (s_1, s_3) =
    \frac{4 \pi^2 \, s_1^3 \, {\cal T}_{\rm GW}
    (s_1, s_3)}
    {(\lambda_* {\cal H}_*)^{2} } \,
    \frac{p_\Pi (s_1)}{s_{\rm turb}\, p_\Pi (s_{\rm turb})},
    \label{S_turb}
    \end{equation}
    where $s_3= \delta t_{fin} f$ and 
\begin{align}
    {\cal T}_{\rm GW} (s_1, s_3) =
    \left\{\begin{array}{ll}
        \ln^2 \bigl[1 + {\cal H}_*
        \delta t_{\rm fin}/(2 \pi) \bigr], & \hspace{-1mm}
        \text{if \ } s_3 < 1, \\
        \ln^2 \bigl[1 + \lambda_* {\cal H}_*/(2\pi s_1)\bigr], &  \hspace{-1mm}
        \text{if \ } s_3 \geq 1.
    \end{array} \right.
\end{align}

    \item $p_\Pi$ is a broken power law based on the von Kárman spectral shape at low frequencies \cite{Durrer_2003} and Kolmogorov spectrum at high frequencies \cite{Kolmogorov}:
    \begin{equation}
    p_\Pi (s_1) \simeq \biggl[1 + \biggl(\frac{s_1}
    {s_\Pi}
    \biggr)^{\alpha_\Pi} \biggr]^{-{11\over{3\alpha_\Pi}}},
    \label{p_pi_fit}
    \end{equation}
where $\alpha_\Pi \simeq 2.15$ and $s_\Pi = 2.2$.
\end{itemize}

\section{Observational constraints}

\subsection{PTA data}

We compare the parameterized SGWB power spectra  described above with the mesaurements by different PTAs.  The PTA results are conventionally presented in the form of the measurements of the normalisation $A_{GW}$ and slope $\gamma$ of the  power spectrum of GW strain
\begin{equation}
h_c(f)=A_{GW}\left(\frac{f}{f_{ref}}\right)^{(3-\gamma)/2}
\end{equation}
 where $f_{ref}=1 yr^{-1}$,

in the frequency range approximately between $10^{-8.75}$~Hz and  $10^{-7.75}$~Hz. The slope of the strain spectrum is related to the slope of the $\Omega_{GW}(f)\propto f^{p}$ as  $p = 5 - \gamma$
\begin{equation}
    \Omega_{GW}(f)= \frac{2 \pi^{2}}{3 H_0^{2}}f^{2}h_c^{2}(f)
\end{equation}
where $H_0$ is the Hubble rate today.
The amplitude $A_{GW}$ is related to the normalisation of $\Omega_{GW}$ as $A_{GW}^{2}= \frac{3H_0^{2}\Omega_{GW}}{2 \pi^{2}f_{ref}^{2}}$. 

For each combination of ($\alpha$, $\beta$, $T$, $v_w$, $\epsilon_{turb}$)  we compute the GW spectrum $\Omega_{GW}=\Omega_{GW}^{sw}+\Omega_{GW}^{turb}$ and measure its slope and normalisation in the same frequency range as for the PTA spectrum $(10\mbox{ yr})^{-1}<f<(2\mbox{ yr})^{-1}$. We compute the normalization $A_{GW}$ and slope $\gamma$ of the strain to compare these parameters to the PTA measurements \cite{NANOGrav:2023gor,EPTA:2023fyk,Reardon:2023gzh}. We derive the $1\sigma$ confidence region in the ($\alpha$, $\beta$, $T$, $v_w$, $\epsilon_{turb}$) parameter space by considering model spectra that have  $\Omega_{GW}(f)$ between the minimal and maximal value allowed by the $1\sigma$ confidence regions in $A_{GW}, \gamma$ parameter space in the specified frequency range.

The PTA results are known to favour "strong" phase transitions with $\alpha\gtrsim 1$ and $v_w\simeq 1$ \cite{NANOGrav:2023hvm,Ellis:2023oxs}. Hence, for the presentation of our results we limit the consideration to $v_w=1$. The value of $\epsilon_{turb}$ is not directly constrained by the PTA data and we arbitrarily fix it to $\epsilon_{turb}=0.1$. For other parameters, we perform calculations in wide intervals $\alpha \in [10^{-1},10^{2}]$,  $\beta \in [1,10^{2}]$, $T \in [10^{-3},10^{1}]$~GeV.

\subsection{Primordial black holes}

A first order phase transition can generate Primordial Black Holes (PBH) \cite{PBH,gouttenoire2023primordial,Kawana:2022olo}. This happens in the case of a strong phase transition $\alpha\gtrsim 1$, if  large enough domains of the old phase ("false vacuum") in which the energy density remains constant $\rho_{old}\sim const$ survive for long time. 
The energy density in the true vacuum region decreases with the growth of the scale factor $a$ as the $\rho_{new} \propto a^{-4}$. The over-density ($\delta\rho=\rho_{old}-\rho_{new}$) creates an excess mass $M\sim R^3\delta\rho$ ($R$ is the size of the old phase domain) and becomes unstable if its size becomes of the order of the gravitational radius, $G_NM\sim R$.

The abundance of the black holes formed through this mechanism, $f_{PBH}$, can be derived from the probability $P(R,t)$ of finding a region of the false vacuum of the size $R$ at the time $t$.
This abundance is naturally constrained to be $f_{PBH}<1$ (or $f_{PBH}\ll 1$ if known observational constraints on the PBH abundance in the relevant mass range are taken into account).
A calculation of the PBH yield for  the case $\alpha\gg 1$, with an additional assumption on the shape and temperature dependence of the potential barrier separating the old and new phase has been done in Ref. \cite{PBH, gouttenoire2023primordial}, which has derived a constraint $\beta \gtrsim 6$ on the phase transition that does not lead to over-production of the PBH.  
The case of moderate $\alpha$ has been considered in Ref.  \cite{gouttenoire2023primordial,Kawana:2022olo} that has found that the lower bound on $\beta$ is $\alpha$-dependent in this situation. In our analysis we adopt the bound from Ref. \cite{Kawana:2022olo}.

\section{Results}

\subsection{PTA and PBH constraints based on new SGWB parameterisations}

The PTA and PBH constraints on the phase transition parameter space $(\alpha, \beta, T_*)$ for fixed values of the bubble wall velocity $v_w\simeq 1$ and turbulence efficiency $\epsilon_{turb}=0.1$ define a region of allowed phase transition parameters in a three-dimensional parameter space. This region can be visualized through its projections on the two-dimensional planes $(\alpha,\beta)$, $(\alpha,T_*)$, $(\beta,T_*)$.

\begin{figure*}
\includegraphics[width=\linewidth]{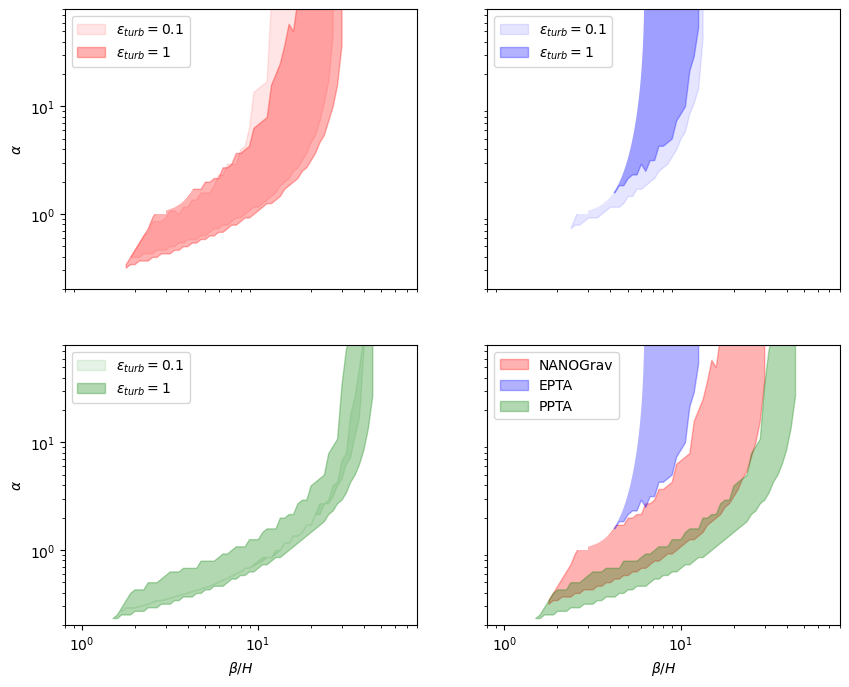}
\caption{$1\sigma$ constraints on $\alpha, \beta$ from the NANOGrav (top left), EPTA (top right) and PPTA (bottom left). Lighter and darker shading shows results for different values of $\epsilon_{turb}$ parameter. Bottom right panel shows an aggregated view of the confidence regions for EPTA (blue), NANOGrav (red) and PPTA (green). } 
\label{alpha_beta}
\end{figure*}

Fig. \ref{alpha_beta} shows the projection of the allowed parameter regions onto the $(\alpha,\beta)$ plane. Different degrees of shading correspond to the different choices of $\epsilon_{turb}$. A point $(\alpha,\beta)$ is considered to be within the allowed parameter range if there is at least one value of $T_*$ for which the SGWB model prediction for $A_{GW}$, $\gamma$ falls within the $1\sigma$ confidence interval of PTA measurements. The PBH constraint cuts a part of the allowed parameter space splitting the $1\sigma$ regions into two "islands" in the case of EPTA  and largely reducing the allowed parameter space for the NANOGrav and PPTA. One can see that the latent heat parameter $\alpha$ is not limited from above. 

\begin{figure*}
\includegraphics[width=\linewidth]{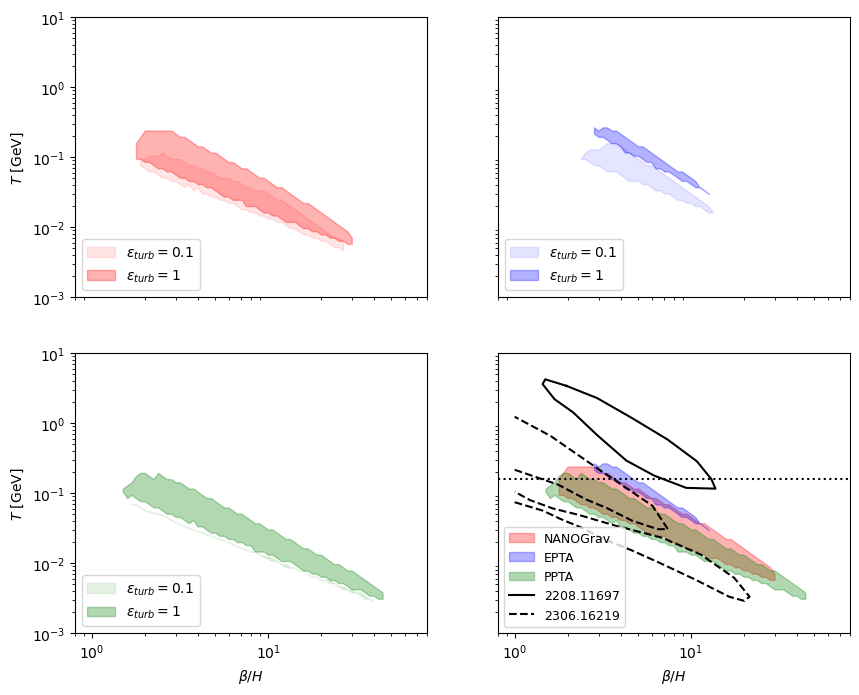}
\caption{Same as in Fig. \ref{alpha_beta} but in projection on the $\beta$, $T_*$  Bottom right panel also shows results reported in Ref. \cite{Ellis:2023oxs} (solid black line) and \cite{EPTA:2023xxk} (two black dashed lines for two different phase transition models). Horizontal line in the last panel shows an estimate of the temperature of the quark confinement \cite{Schwarz:2003du}. } 
\label{beta_T}
\end{figure*}

Figure \ref{beta_T} shows the projection of the allowed parameter space onto $\beta-T_*$ plane. One can see that the allowed region is elongated along certain direction, favoring the models in which the characteristic frequency corresponding to the phase transition is close to the observed frequency range.

\begin{figure*}
\includegraphics[width=\textwidth]{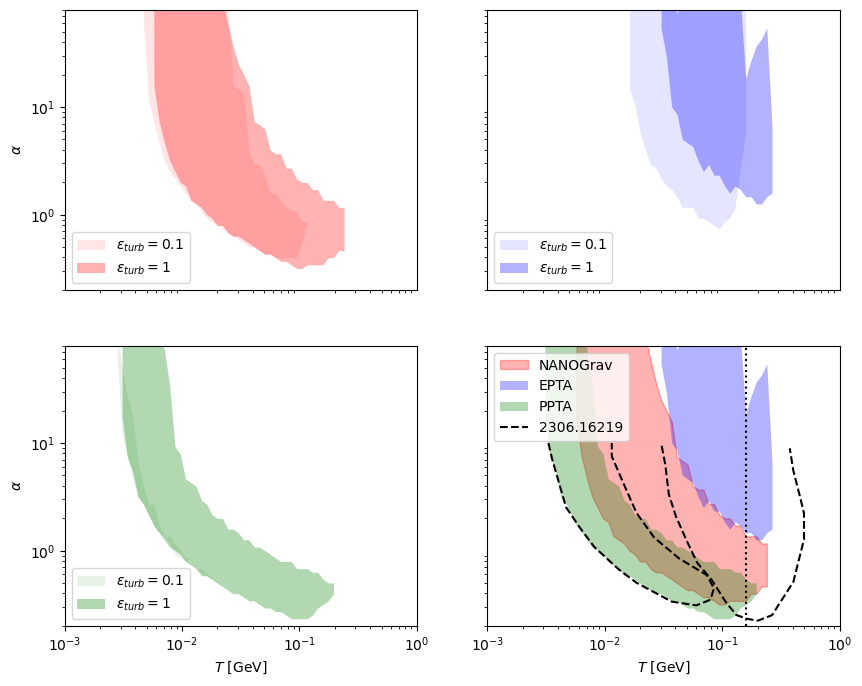}
\caption{Same as in Fig. \ref{beta_T}, but for the $(\alpha, T_*)$ projection. } 
\label{alpha_T}
\end{figure*}

Fig. \ref{alpha_T} show the projection of the allowed parameter space  on $\alpha-T_*$ plane. Also here the PBH constraint separates the $1\sigma$ confidence region into two separate regions for the EPTA. 

Bottom right panels of Figs. \ref{alpha_beta}, \ref{beta_T}, \ref{alpha_T} show comparison of the $1\sigma$ confidence contours for three different PTAs.   

\subsection{Comparison with previous analysis }

\begin{figure}
    \includegraphics[width=\columnwidth]{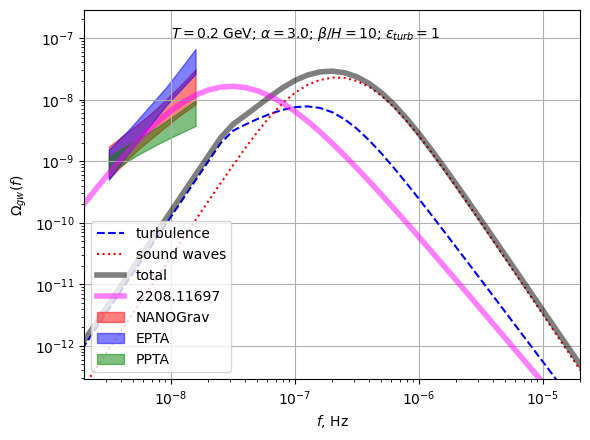}
    \caption{Comparison of the model SGWB power spectra for the parameterisations of Ref. \cite{Lewicki:2022pdb,Ellis:2023oxs} (magenta line) and of Ref. \cite{RoperPol:2023bqa} (red dashed, blue dotted and grey solid lines) for the same phase transition parameters ($T_*=0.2$~GeV, $\alpha=3.0, \beta=10H(T_*),v_w=1.0, \epsilon_{turb}=1$. Red, blue and green "butterflies" show the NANOGrav, EPTA and PPTA measurements of the SGWB spectrum\cite{NANOGrav:2023gor,EPTA:2023fyk,Reardon:2023gzh}. Calculation of the powerspectrum is available as a service at \protect\url{https://mmoda.io} and \protect\url{https://usegalaxy.eu} \cite{service}.}
    \label{spectrum_diff}
\end{figure}

Parameterisations of the gravitational wave spectrum used above differ from those of  Refs. \cite{NANOGrav:2023hvm,Ellis:2023oxs}. Bottom right panels of Fig. \ref{beta_T}, \ref{alpha_T} show the difference in the parameter estimates  resulting from this difference in parameterisation.  Similar to our analysis, the analysis of Ref. \cite{NANOGrav:2023hvm} is applicable to moderate $\alpha\lesssim 1$, while the analysis of \cite{Ellis:2023oxs} is  limited to $\alpha\gg 1$. As a result, our results generally agree with those of \cite{NANOGrav:2023hvm} for $\beta, T_*$ (Fig. \ref{beta_T}) but reveal a substantial difference with the analysis of \cite{Ellis:2023oxs}. 

Fig. \ref{spectrum_diff} shows that difference between the shape of the gravitational wave power spectrum for the same values of the phase transition parameters for the parameterisations used in our analysis and those of \cite{Ellis:2023oxs}. One can see that even though the overall shapes of the spectrum are similar, the peak of the spectrum is shifted toward lower frequencies in the \cite{Ellis:2023oxs} parameterisation, to $2\pi f\sim 0.7 H_* $ where $H_*$ is the Hubble rate at the moment of the phase transition, while the modelling of \cite{Hindmarsh_2019}, that serves as a basis for the parameterisations used in our analysis, is at  $2\pi f\sim 4 H_* $. The PTA data are consistent with a possibility $\alpha\gtrsim 1$ so that it is not clear a-priori, which of the two approximations, $\alpha\lesssim 1$ or $\alpha\gg 1$ is more suitable for modelling of the data. 

\section{Implications for the cosmological magnetic field}

\begin{figure}
\includegraphics[width=\columnwidth]{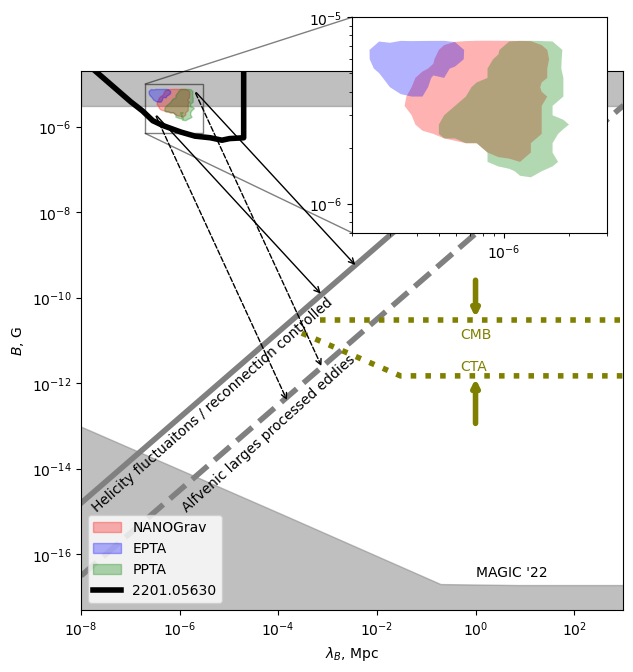}
\caption{Estimates of the co-moving strength and correlation length of the cosmological magnetic field at the moment of generation from the NANOGrav (red), EPTA (blue) and PPTA (green) data, based on the model of Ref. \cite{RoperPol:2023bqa}. Black thick line shows the estimate from previously published NANOGrav "hint" of the SGWB \cite{pol2022gravitational}. Inclined arrows show possible evolutionary tracks of $\lambda_B, B$, according to Refs. \cite{Hosking:2022umv} (solid) and \cite{Banerjee:2004df} (dashed). Grey inclined lines show possible loci of the evolution endpoints, from Refs. \cite{Banerjee:2004df} (dashed) and \cite{Brandenburg:2024tyi} (solid). Lower bound on the present-day strength of IGMF, shown in grey, is from Ref. \cite{MAGIC:2022piy}. The grey  upper bound is at the level $3\ \mu$G, a field for which the energy density is in equipartition with the the Cosmic Microwave Backgorund (CMB). Dotted olive line with upward arrow  is the sensitivity reach of CTA \cite{Korochkin:2020pvg}. Dotted olive line with downward arrow is the sensitivity reach of CMB measurements \cite{Galli:2021mxk}. $\epsilon_{turb}=1$ is assumed for the estimate. }
  \label{B_results}
\end{figure}

The turbulent plasma motions excited by the first-order phase transition result in production of magnetic field that can survive till the present day in the form of the Inter-Galactic Magnetic Field (IGMF) \cite{Durrer:2013pga}. The cosmological IGMF may reside in the present-day voids of the Large Scale Structure (LSS). Results from $\gamma$-ray telescopes indicate that the voids of the LSS do have non-zero IGMF \cite{2010Sci...328...73N,MAGIC:2022piy}. The void IGMF gets amplified in the filaments of the LSS where it can be traced by the observations of the Faraday rotation of radio signals from distant extragalactic sources. Recent LOFAR telescope observations provide a measurement of the magnetic field in the filaments and possibly provide an indication that this field is of cosmological origin \cite{Carretti:2022fqk}.  

The strength $B$ and correlation length $\lambda_B$ of the cosmological magnetic field found in the voids are related to its initial strength $B_*$ and correlation length $\lambda_*$ through transfer functions that are calculated based on the MHD modelling of the field evolution throughout cosmological epochs \cite{Banerjee:2004df,Hosking:2022umv}. The PTA observations constrain the initial configuration, $B_*,\lambda_*$. The initial correlation length of the magnetic field is determined by the bubble's radius at percolation $\lambda_*$, given by Eq. (\ref{lambda}). The strength of the field is determined by the fraction of the energy output of the phase transition that is transferred to the MHD turbulence. We adopt the approach of Ref. \cite{RoperPol:2023bqa} for the estimation of the magnetic field parameters $B_*, \lambda_*$. 

Figure \ref{B_results} shows the constraints on the $B_*,\lambda_*$ derived from the new PTA data. These constraints are much tighter than previously derived constraints based on a shorter NANOGrav data \cite{Roper_Pol_2022}. Most remarkable improvement is shrinking of the range of allowed correlation lengths of magnetic field.  One can see that the uncertainty range of $\lambda_*$ is reduced to a factor of 3, the co-moving correlation length is bound to be close to one co-moving parsec (compared to the broad range $0.1$~pc$<\lambda_*<30$~pc for the previous analysis). This tighter constraint is due to the alignment of the constraints in the $T_*-\beta$ projection along the direciton that corresponds to the constant size of the new phase bubbles at the onset of percolation.  

Adopting an estimate $\epsilon_{turb}\sim 0.1..1$, one also finds a rather precise estimate of the strength of the magnetic field, $B\sim 1-3 \mu$G, roughly in equipartition with photon energy density at the moment of production for $\epsilon_{turb}=1$. 

The expected evolution of the field with initial parameters $B_*,\lambda_*$ is shown by the black solid and dashed arrows for two possible turbulence decay laws of non-helical magnetic fields \cite{Banerjee:2004df,Hosking:2022umv}. We also consider different possibilities for the endpoints of evolution: the "Alfvenic" largest processed eddies scale from Ref. \cite{Banerjee:2004df} and the scales suggested by the numerical modelling of the evolution \cite{Hosking:2022umv,Brandenburg:2024tyi}. If the turbulent decay proceeds according to the molde of \cite{Hosking:2022umv,Brandenburg:2024tyi}, the phase transition relic IGMF occupying the voids of the Large Scale Structure today has the strength about $10^{-10}$~G and correlation length about 1~kpc. Such field should have affected the CMB observables \cite{Galli:2021mxk} and possibly also the structure formation \cite{2020A&A...643A..54S} and its imprint on the CMB and LSS may be detectable. This field is above the sensitivity reach of the $\gamma$-ray technique (the line with "CTA" tag in Fig. \ref{B_results}). To the contrary, if the field has evolved according to the model of Ref. \cite{Banerjee:2004df}, it is measurable with the next generation gamma-ray observatory Cherenkov Telescope Array (CTA) \cite{Korochkin:2020pvg}. In any case the present-day relic field with initial parameters derived from the PTA data is measurable at later cosmological epochs and hence has a "multi-messenger" signature.

\section{Conclusions}

The results derived above provide an update and an improvement on previously reported estimates of the parameters of the cosmological phase transition and cosmological magnetic field consistent with the PTA data. They also provide an aggregated view of the constraints on the cosmological phase transition parameters extracted from three different PTA datasets,    EPTA, PPTA and NANOGrav. Compared to the analysis of Refs. \cite{Roper_Pol_2022,EPTA:2023xxk} that have concentrated on the effect on the MHD turbulence, our analysis considers two-component gravitational wave spectrum: from sound-waves generated at percolation between bubbles and from the MHD turbulence. The analysis also includes the constraint from the PBH over-production. 

\bibliography{refs}

\begin{thebibliography}{43}
\expandafter\ifx\csname natexlab\endcsname\relax\def\natexlab#1{#1}\fi
\expandafter\ifx\csname bibnamefont\endcsname\relax
  \def\bibnamefont#1{#1}\fi
\expandafter\ifx\csname bibfnamefont\endcsname\relax
  \def\bibfnamefont#1{#1}\fi
\expandafter\ifx\csname citenamefont\endcsname\relax
  \def\citenamefont#1{#1}\fi
\expandafter\ifx\csname url\endcsname\relax
  \def\url#1{\texttt{#1}}\fi
\expandafter\ifx\csname urlprefix\endcsname\relax\def\urlprefix{URL }\fi
\providecommand{\bibinfo}[2]{#2}
\providecommand{\eprint}[2][]{\url{#2}}

\bibitem[{\citenamefont{Agazie et~al.}(2023)}]{NANOGrav:2023gor}
\bibinfo{author}{\bibfnamefont{G.}~\bibnamefont{Agazie}} \bibnamefont{et~al.}
  (\bibinfo{collaboration}{NANOGrav}), \bibinfo{journal}{Astrophys. J. Lett.}
  \textbf{\bibinfo{volume}{951}}, \bibinfo{pages}{L8} (\bibinfo{year}{2023}),
  \eprint{2306.16213}.

\bibitem[{\citenamefont{Antoniadis et~al.}(2023{\natexlab{a}})}]{EPTA:2023fyk}
\bibinfo{author}{\bibfnamefont{J.}~\bibnamefont{Antoniadis}}
  \bibnamefont{et~al.} (\bibinfo{collaboration}{EPTA})
  (\bibinfo{year}{2023}{\natexlab{a}}), \eprint{2306.16214}.

\bibitem[{\citenamefont{Reardon et~al.}(2023)}]{Reardon:2023gzh}
\bibinfo{author}{\bibfnamefont{D.~J.} \bibnamefont{Reardon}}
  \bibnamefont{et~al.}, \bibinfo{journal}{Astrophys. J. Lett.}
  \textbf{\bibinfo{volume}{951}}, \bibinfo{pages}{L6} (\bibinfo{year}{2023}),
  \eprint{2306.16215}.

\bibitem[{\citenamefont{Xu et~al.}(2023)}]{Xu:2023wog}
\bibinfo{author}{\bibfnamefont{H.}~\bibnamefont{Xu}} \bibnamefont{et~al.},
  \bibinfo{journal}{Res. Astron. Astrophys.} \textbf{\bibinfo{volume}{23}},
  \bibinfo{pages}{075024} (\bibinfo{year}{2023}), \eprint{2306.16216}.

\bibitem[{\citenamefont{Burke-Spolaor et~al.}(2019)}]{Burke-Spolaor:2018bvk}
\bibinfo{author}{\bibfnamefont{S.}~\bibnamefont{Burke-Spolaor}}
  \bibnamefont{et~al.}, \bibinfo{journal}{Astron. Astrophys. Rev.}
  \textbf{\bibinfo{volume}{27}}, \bibinfo{pages}{5} (\bibinfo{year}{2019}),
  \eprint{1811.08826}.

\bibitem[{\citenamefont{Sato-Polito et~al.}(2023)\citenamefont{Sato-Polito,
  Zaldarriaga, and Quataert}}]{Sato-Polito:2023gym}
\bibinfo{author}{\bibfnamefont{G.}~\bibnamefont{Sato-Polito}},
  \bibinfo{author}{\bibfnamefont{M.}~\bibnamefont{Zaldarriaga}},
  \bibnamefont{and} \bibinfo{author}{\bibfnamefont{E.}~\bibnamefont{Quataert}}
  (\bibinfo{year}{2023}), \eprint{2312.06756}.

\bibitem[{\citenamefont{Afzal et~al.}(2023)}]{NANOGrav:2023hvm}
\bibinfo{author}{\bibfnamefont{A.}~\bibnamefont{Afzal}} \bibnamefont{et~al.}
  (\bibinfo{collaboration}{NANOGrav}), \bibinfo{journal}{Astrophys. J. Lett.}
  \textbf{\bibinfo{volume}{951}}, \bibinfo{pages}{L11} (\bibinfo{year}{2023}),
  \eprint{2306.16219}.

\bibitem[{\citenamefont{Witten}(1984)}]{PhysRevD.30.272}
\bibinfo{author}{\bibfnamefont{E.}~\bibnamefont{Witten}},
  \bibinfo{journal}{Phys. Rev. D} \textbf{\bibinfo{volume}{30}},
  \bibinfo{pages}{272} (\bibinfo{year}{1984}),
  \urlprefix\url{https://link.aps.org/doi/10.1103/PhysRevD.30.272}.

\bibitem[{\citenamefont{Kosowsky and Turner}(1993)}]{Kosowsky:1992vn}
\bibinfo{author}{\bibfnamefont{A.}~\bibnamefont{Kosowsky}} \bibnamefont{and}
  \bibinfo{author}{\bibfnamefont{M.~S.} \bibnamefont{Turner}},
  \bibinfo{journal}{Phys. Rev. D} \textbf{\bibinfo{volume}{47}},
  \bibinfo{pages}{4372} (\bibinfo{year}{1993}), \eprint{astro-ph/9211004}.

\bibitem[{\citenamefont{Caprini et~al.}(2008)\citenamefont{Caprini, Durrer, and
  Servant}}]{Caprini:2007xq}
\bibinfo{author}{\bibfnamefont{C.}~\bibnamefont{Caprini}},
  \bibinfo{author}{\bibfnamefont{R.}~\bibnamefont{Durrer}}, \bibnamefont{and}
  \bibinfo{author}{\bibfnamefont{G.}~\bibnamefont{Servant}},
  \bibinfo{journal}{Phys. Rev. D} \textbf{\bibinfo{volume}{77}},
  \bibinfo{pages}{124015} (\bibinfo{year}{2008}), \eprint{0711.2593}.

\bibitem[{\citenamefont{Arzoumanian et~al.}(2021)}]{NANOGrav:2021flc}
\bibinfo{author}{\bibfnamefont{Z.}~\bibnamefont{Arzoumanian}}
  \bibnamefont{et~al.} (\bibinfo{collaboration}{NANOGrav}),
  \bibinfo{journal}{Phys. Rev. Lett.} \textbf{\bibinfo{volume}{127}},
  \bibinfo{pages}{251302} (\bibinfo{year}{2021}), \eprint{2104.13930}.

\bibitem[{\citenamefont{Sigl et~al.}(1997)\citenamefont{Sigl, Olinto, and
  Jedamzik}}]{Sigl:1996dm}
\bibinfo{author}{\bibfnamefont{G.}~\bibnamefont{Sigl}},
  \bibinfo{author}{\bibfnamefont{A.~V.} \bibnamefont{Olinto}},
  \bibnamefont{and} \bibinfo{author}{\bibfnamefont{K.}~\bibnamefont{Jedamzik}},
  \bibinfo{journal}{Phys. Rev. D} \textbf{\bibinfo{volume}{55}},
  \bibinfo{pages}{4582} (\bibinfo{year}{1997}), \eprint{astro-ph/9610201}.

\bibitem[{\citenamefont{Neronov et~al.}(2021)\citenamefont{Neronov, Roper~Pol,
  Caprini, and Semikoz}}]{Neronov:2020qrl}
\bibinfo{author}{\bibfnamefont{A.}~\bibnamefont{Neronov}},
  \bibinfo{author}{\bibfnamefont{A.}~\bibnamefont{Roper~Pol}},
  \bibinfo{author}{\bibfnamefont{C.}~\bibnamefont{Caprini}}, \bibnamefont{and}
  \bibinfo{author}{\bibfnamefont{D.}~\bibnamefont{Semikoz}},
  \bibinfo{journal}{Phys. Rev. D} \textbf{\bibinfo{volume}{103}},
  \bibinfo{pages}{041302} (\bibinfo{year}{2021}), \eprint{2009.14174}.

\bibitem[{\citenamefont{Roper~Pol
  et~al.}(2022{\natexlab{a}})\citenamefont{Roper~Pol, Caprini, Neronov, and
  Semikoz}}]{RoperPol:2022iel}
\bibinfo{author}{\bibfnamefont{A.}~\bibnamefont{Roper~Pol}},
  \bibinfo{author}{\bibfnamefont{C.}~\bibnamefont{Caprini}},
  \bibinfo{author}{\bibfnamefont{A.}~\bibnamefont{Neronov}}, \bibnamefont{and}
  \bibinfo{author}{\bibfnamefont{D.}~\bibnamefont{Semikoz}},
  \bibinfo{journal}{Phys. Rev. D} \textbf{\bibinfo{volume}{105}},
  \bibinfo{pages}{123502} (\bibinfo{year}{2022}{\natexlab{a}}),
  \eprint{2201.05630}.

\bibitem[{\citenamefont{Brandenburg et~al.}(2021)\citenamefont{Brandenburg,
  Clarke, He, and Kahniashvili}}]{Brandenburg:2021tmp}
\bibinfo{author}{\bibfnamefont{A.}~\bibnamefont{Brandenburg}},
  \bibinfo{author}{\bibfnamefont{E.}~\bibnamefont{Clarke}},
  \bibinfo{author}{\bibfnamefont{Y.}~\bibnamefont{He}}, \bibnamefont{and}
  \bibinfo{author}{\bibfnamefont{T.}~\bibnamefont{Kahniashvili}},
  \bibinfo{journal}{Phys. Rev. D} \textbf{\bibinfo{volume}{104}},
  \bibinfo{pages}{043513} (\bibinfo{year}{2021}), \eprint{2102.12428}.

\bibitem[{\citenamefont{Antoniadis et~al.}(2023{\natexlab{b}})}]{EPTA:2023xxk}
\bibinfo{author}{\bibfnamefont{J.}~\bibnamefont{Antoniadis}}
  \bibnamefont{et~al.} (\bibinfo{collaboration}{EPTA})
  (\bibinfo{year}{2023}{\natexlab{b}}), \eprint{2306.16227}.

\bibitem[{\citenamefont{Jedamzik and Pogosian}(2020)}]{Jedamzik:2020krr}
\bibinfo{author}{\bibfnamefont{K.}~\bibnamefont{Jedamzik}} \bibnamefont{and}
  \bibinfo{author}{\bibfnamefont{L.}~\bibnamefont{Pogosian}},
  \bibinfo{journal}{Phys. Rev. Lett.} \textbf{\bibinfo{volume}{125}},
  \bibinfo{pages}{181302} (\bibinfo{year}{2020}), \eprint{2004.09487}.

\bibitem[{\citenamefont{Galli et~al.}(2022)\citenamefont{Galli, Pogosian,
  Jedamzik, and Balkenhol}}]{Galli:2021mxk}
\bibinfo{author}{\bibfnamefont{S.}~\bibnamefont{Galli}},
  \bibinfo{author}{\bibfnamefont{L.}~\bibnamefont{Pogosian}},
  \bibinfo{author}{\bibfnamefont{K.}~\bibnamefont{Jedamzik}}, \bibnamefont{and}
  \bibinfo{author}{\bibfnamefont{L.}~\bibnamefont{Balkenhol}},
  \bibinfo{journal}{Phys. Rev. D} \textbf{\bibinfo{volume}{105}},
  \bibinfo{pages}{023513} (\bibinfo{year}{2022}), \eprint{2109.03816}.

\bibitem[{\citenamefont{Ellis et~al.}(2023)\citenamefont{Ellis, Fairbairn,
  Franciolini, H\"utsi, Iovino, Lewicki, Raidal, Urrutia, Vaskonen, and
  Veerm\"ae}}]{Ellis:2023oxs}
\bibinfo{author}{\bibfnamefont{J.}~\bibnamefont{Ellis}},
  \bibinfo{author}{\bibfnamefont{M.}~\bibnamefont{Fairbairn}},
  \bibinfo{author}{\bibfnamefont{G.}~\bibnamefont{Franciolini}},
  \bibinfo{author}{\bibfnamefont{G.}~\bibnamefont{H\"utsi}},
  \bibinfo{author}{\bibfnamefont{A.}~\bibnamefont{Iovino}},
  \bibinfo{author}{\bibfnamefont{M.}~\bibnamefont{Lewicki}},
  \bibinfo{author}{\bibfnamefont{M.}~\bibnamefont{Raidal}},
  \bibinfo{author}{\bibfnamefont{J.}~\bibnamefont{Urrutia}},
  \bibinfo{author}{\bibfnamefont{V.}~\bibnamefont{Vaskonen}}, \bibnamefont{and}
  \bibinfo{author}{\bibfnamefont{H.}~\bibnamefont{Veerm\"ae}}
  (\bibinfo{year}{2023}), \eprint{2308.08546}.

\bibitem[{\citenamefont{Roper~Pol et~al.}(2023)\citenamefont{Roper~Pol,
  Neronov, Caprini, Boyer, and Semikoz}}]{RoperPol:2023bqa}
\bibinfo{author}{\bibfnamefont{A.}~\bibnamefont{Roper~Pol}},
  \bibinfo{author}{\bibfnamefont{A.}~\bibnamefont{Neronov}},
  \bibinfo{author}{\bibfnamefont{C.}~\bibnamefont{Caprini}},
  \bibinfo{author}{\bibfnamefont{T.}~\bibnamefont{Boyer}}, \bibnamefont{and}
  \bibinfo{author}{\bibfnamefont{D.}~\bibnamefont{Semikoz}}
  (\bibinfo{year}{2023}), \eprint{2307.10744}.

\bibitem[{\citenamefont{Hindmarsh and Hijazi}(2019)}]{Hindmarsh_2019}
\bibinfo{author}{\bibfnamefont{M.}~\bibnamefont{Hindmarsh}} \bibnamefont{and}
  \bibinfo{author}{\bibfnamefont{M.}~\bibnamefont{Hijazi}},
  \bibinfo{journal}{Journal of Cosmology and Astroparticle Physics}
  \textbf{\bibinfo{volume}{2019}}, \bibinfo{pages}{062–062}
  (\bibinfo{year}{2019}), ISSN \bibinfo{issn}{1475-7516},
  \urlprefix\url{http://dx.doi.org/10.1088/1475-7516/2019/12/062}.

\bibitem[{\citenamefont{Hindmarsh et~al.}(2017)\citenamefont{Hindmarsh, Huber,
  Rummukainen, and Weir}}]{B(delta_w)}
\bibinfo{author}{\bibfnamefont{M.}~\bibnamefont{Hindmarsh}},
  \bibinfo{author}{\bibfnamefont{S.~J.} \bibnamefont{Huber}},
  \bibinfo{author}{\bibfnamefont{K.}~\bibnamefont{Rummukainen}},
  \bibnamefont{and} \bibinfo{author}{\bibfnamefont{D.~J.} \bibnamefont{Weir}}
  (\bibinfo{year}{2017}), \eprint{1704.05871}.

\bibitem[{\citenamefont{Espinosa et~al.}(2010)\citenamefont{Espinosa,
  Konstandin, No, and Servant}}]{Espinosa_2010}
\bibinfo{author}{\bibfnamefont{J.~R.} \bibnamefont{Espinosa}},
  \bibinfo{author}{\bibfnamefont{T.}~\bibnamefont{Konstandin}},
  \bibinfo{author}{\bibfnamefont{J.~M.} \bibnamefont{No}}, \bibnamefont{and}
  \bibinfo{author}{\bibfnamefont{G.}~\bibnamefont{Servant}},
  \bibinfo{journal}{Journal of Cosmology and Astroparticle Physics}
  \textbf{\bibinfo{volume}{2010}}, \bibinfo{pages}{028–028}
  (\bibinfo{year}{2010}), ISSN \bibinfo{issn}{1475-7516},
  \urlprefix\url{http://dx.doi.org/10.1088/1475-7516/2010/06/028}.

\bibitem[{\citenamefont{Jinno et~al.}(2023)\citenamefont{Jinno, Konstandin,
  Rubira, and Stomberg}}]{SWGW}
\bibinfo{author}{\bibfnamefont{R.}~\bibnamefont{Jinno}},
  \bibinfo{author}{\bibfnamefont{T.}~\bibnamefont{Konstandin}},
  \bibinfo{author}{\bibfnamefont{H.}~\bibnamefont{Rubira}}, \bibnamefont{and}
  \bibinfo{author}{\bibfnamefont{I.}~\bibnamefont{Stomberg}},
  \bibinfo{journal}{Journal of Cosmology and Astroparticle Physics}
  \textbf{\bibinfo{volume}{2023}}, \bibinfo{pages}{011} (\bibinfo{year}{2023}).

\bibitem[{\citenamefont{Roper~Pol
  et~al.}(2022{\natexlab{b}})\citenamefont{Roper~Pol, Caprini, Neronov, and
  Semikoz}}]{Roper_Pol_2022}
\bibinfo{author}{\bibfnamefont{A.}~\bibnamefont{Roper~Pol}},
  \bibinfo{author}{\bibfnamefont{C.}~\bibnamefont{Caprini}},
  \bibinfo{author}{\bibfnamefont{A.}~\bibnamefont{Neronov}}, \bibnamefont{and}
  \bibinfo{author}{\bibfnamefont{D.}~\bibnamefont{Semikoz}},
  \bibinfo{journal}{Physical Review D} \textbf{\bibinfo{volume}{105}}
  (\bibinfo{year}{2022}{\natexlab{b}}), ISSN \bibinfo{issn}{2470-0029},
  \urlprefix\url{http://dx.doi.org/10.1103/PhysRevD.105.123502}.

\bibitem[{\citenamefont{Durrer and Caprini}(2003)}]{Durrer_2003}
\bibinfo{author}{\bibfnamefont{R.}~\bibnamefont{Durrer}} \bibnamefont{and}
  \bibinfo{author}{\bibfnamefont{C.}~\bibnamefont{Caprini}},
  \bibinfo{journal}{Journal of Cosmology and Astroparticle Physics}
  \textbf{\bibinfo{volume}{2003}}, \bibinfo{pages}{010–010}
  (\bibinfo{year}{2003}), ISSN \bibinfo{issn}{1475-7516},
  \urlprefix\url{http://dx.doi.org/10.1088/1475-7516/2003/11/010}.

\bibitem[{\citenamefont{Karman}(1948)}]{Kolmogorov}
\bibinfo{author}{\bibfnamefont{T.~V.} \bibnamefont{Karman}}
  (\bibinfo{year}{1948}).

\bibitem[{\citenamefont{Lewicki et~al.}(2023)\citenamefont{Lewicki, Toczek, and
  Vaskonen}}]{PBH}
\bibinfo{author}{\bibfnamefont{M.}~\bibnamefont{Lewicki}},
  \bibinfo{author}{\bibfnamefont{P.}~\bibnamefont{Toczek}}, \bibnamefont{and}
  \bibinfo{author}{\bibfnamefont{V.}~\bibnamefont{Vaskonen}},
  \bibinfo{journal}{Journal of High Energy Physics}
  \textbf{\bibinfo{volume}{2023}} (\bibinfo{year}{2023}), ISSN
  \bibinfo{issn}{1029-8479},
  \urlprefix\url{http://dx.doi.org/10.1007/JHEP09(2023)092}.

\bibitem[{\citenamefont{Gouttenoire and
  Volansky}(2023)}]{gouttenoire2023primordial}
\bibinfo{author}{\bibfnamefont{Y.}~\bibnamefont{Gouttenoire}} \bibnamefont{and}
  \bibinfo{author}{\bibfnamefont{T.}~\bibnamefont{Volansky}},
  \emph{\bibinfo{title}{Primordial black holes from supercooled phase
  transitions}} (\bibinfo{year}{2023}), \eprint{2305.04942}.

\bibitem[{\citenamefont{Kawana et~al.}(2023)\citenamefont{Kawana, Kim, and
  Lu}}]{Kawana:2022olo}
\bibinfo{author}{\bibfnamefont{K.}~\bibnamefont{Kawana}},
  \bibinfo{author}{\bibfnamefont{T.}~\bibnamefont{Kim}}, \bibnamefont{and}
  \bibinfo{author}{\bibfnamefont{P.}~\bibnamefont{Lu}}, \bibinfo{journal}{Phys.
  Rev. D} \textbf{\bibinfo{volume}{108}}, \bibinfo{pages}{103531}
  (\bibinfo{year}{2023}), \eprint{2212.14037}.

\bibitem[{\citenamefont{Schwarz}(2003)}]{Schwarz:2003du}
\bibinfo{author}{\bibfnamefont{D.~J.} \bibnamefont{Schwarz}},
  \bibinfo{journal}{Annalen Phys.} \textbf{\bibinfo{volume}{12}},
  \bibinfo{pages}{220} (\bibinfo{year}{2003}), \eprint{astro-ph/0303574}.

\bibitem[{\citenamefont{Lewicki and Vaskonen}(2023)}]{Lewicki:2022pdb}
\bibinfo{author}{\bibfnamefont{M.}~\bibnamefont{Lewicki}} \bibnamefont{and}
  \bibinfo{author}{\bibfnamefont{V.}~\bibnamefont{Vaskonen}},
  \bibinfo{journal}{Eur. Phys. J. C} \textbf{\bibinfo{volume}{83}},
  \bibinfo{pages}{109} (\bibinfo{year}{2023}), \eprint{2208.11697}.

\bibitem[{\citenamefont{Neronov and Boyer}(2024)}]{service}
\bibinfo{author}{\bibfnamefont{A.}~\bibnamefont{Neronov}} \bibnamefont{and}
  \bibinfo{author}{\bibfnamefont{T.}~\bibnamefont{Boyer}},
  \bibinfo{journal}{WorkflowHub}
  \bibinfo{eid}{https://doi.org/10.48546/workflowhub.workflow.831.1}
  (\bibinfo{year}{2024}).

\bibitem[{\citenamefont{Pol}(2022)}]{pol2022gravitational}
\bibinfo{author}{\bibfnamefont{A.~R.} \bibnamefont{Pol}},
  \emph{\bibinfo{title}{Gravitational waves from mhd turbulence at the qcd
  phase transition as a source for pulsar timing arrays}}
  (\bibinfo{year}{2022}), \eprint{2205.09261}.

\bibitem[{\citenamefont{Hosking and Schekochihin}(2023)}]{Hosking:2022umv}
\bibinfo{author}{\bibfnamefont{D.~N.} \bibnamefont{Hosking}} \bibnamefont{and}
  \bibinfo{author}{\bibfnamefont{A.~A.} \bibnamefont{Schekochihin}},
  \bibinfo{journal}{Nature Commun.} \textbf{\bibinfo{volume}{14}},
  \bibinfo{pages}{7523} (\bibinfo{year}{2023}), \eprint{2203.03573}.

\bibitem[{\citenamefont{Banerjee and Jedamzik}(2004)}]{Banerjee:2004df}
\bibinfo{author}{\bibfnamefont{R.}~\bibnamefont{Banerjee}} \bibnamefont{and}
  \bibinfo{author}{\bibfnamefont{K.}~\bibnamefont{Jedamzik}},
  \bibinfo{journal}{Phys. Rev. D} \textbf{\bibinfo{volume}{70}},
  \bibinfo{pages}{123003} (\bibinfo{year}{2004}), \eprint{astro-ph/0410032}.

\bibitem[{\citenamefont{Brandenburg et~al.}(2024)\citenamefont{Brandenburg,
  Neronov, and Vazza}}]{Brandenburg:2024tyi}
\bibinfo{author}{\bibfnamefont{A.}~\bibnamefont{Brandenburg}},
  \bibinfo{author}{\bibfnamefont{A.}~\bibnamefont{Neronov}}, \bibnamefont{and}
  \bibinfo{author}{\bibfnamefont{F.}~\bibnamefont{Vazza}}
  (\bibinfo{year}{2024}), \eprint{2401.08569}.

\bibitem[{\citenamefont{Acciari et~al.}(2023)}]{MAGIC:2022piy}
\bibinfo{author}{\bibfnamefont{V.~A.} \bibnamefont{Acciari}}
  \bibnamefont{et~al.} (\bibinfo{collaboration}{MAGIC}),
  \bibinfo{journal}{Astron. Astrophys.} \textbf{\bibinfo{volume}{670}},
  \bibinfo{pages}{A145} (\bibinfo{year}{2023}), \eprint{2210.03321}.

\bibitem[{\citenamefont{Korochkin et~al.}(2021)\citenamefont{Korochkin,
  Kalashev, Neronov, and Semikoz}}]{Korochkin:2020pvg}
\bibinfo{author}{\bibfnamefont{A.}~\bibnamefont{Korochkin}},
  \bibinfo{author}{\bibfnamefont{O.}~\bibnamefont{Kalashev}},
  \bibinfo{author}{\bibfnamefont{A.}~\bibnamefont{Neronov}}, \bibnamefont{and}
  \bibinfo{author}{\bibfnamefont{D.}~\bibnamefont{Semikoz}},
  \bibinfo{journal}{Astrophys. J.} \textbf{\bibinfo{volume}{906}},
  \bibinfo{pages}{116} (\bibinfo{year}{2021}), \eprint{2007.14331}.

\bibitem[{\citenamefont{Durrer and Neronov}(2013)}]{Durrer:2013pga}
\bibinfo{author}{\bibfnamefont{R.}~\bibnamefont{Durrer}} \bibnamefont{and}
  \bibinfo{author}{\bibfnamefont{A.}~\bibnamefont{Neronov}},
  \bibinfo{journal}{Astron. Astrophys. Rev.} \textbf{\bibinfo{volume}{21}},
  \bibinfo{pages}{62} (\bibinfo{year}{2013}), \eprint{1303.7121}.

\bibitem[{\citenamefont{{Neronov} and {Vovk}}(2010)}]{2010Sci...328...73N}
\bibinfo{author}{\bibfnamefont{A.}~\bibnamefont{{Neronov}}} \bibnamefont{and}
  \bibinfo{author}{\bibfnamefont{I.}~\bibnamefont{{Vovk}}},
  \bibinfo{journal}{Science} \textbf{\bibinfo{volume}{328}},
  \bibinfo{pages}{73} (\bibinfo{year}{2010}), \eprint{1006.3504}.

\bibitem[{\citenamefont{Carretti et~al.}(2022)\citenamefont{Carretti,
  O\textquoteright{}Sullivan, Vacca, Vazza, Gheller, Vernstrom, and
  Bonafede}}]{Carretti:2022fqk}
\bibinfo{author}{\bibfnamefont{E.}~\bibnamefont{Carretti}},
  \bibinfo{author}{\bibfnamefont{S.~P.}
  \bibnamefont{O\textquoteright{}Sullivan}},
  \bibinfo{author}{\bibfnamefont{V.}~\bibnamefont{Vacca}},
  \bibinfo{author}{\bibfnamefont{F.}~\bibnamefont{Vazza}},
  \bibinfo{author}{\bibfnamefont{C.}~\bibnamefont{Gheller}},
  \bibinfo{author}{\bibfnamefont{T.}~\bibnamefont{Vernstrom}},
  \bibnamefont{and} \bibinfo{author}{\bibfnamefont{A.}~\bibnamefont{Bonafede}},
  \bibinfo{journal}{Mon. Not. Roy. Astron. Soc.}
  \textbf{\bibinfo{volume}{518}}, \bibinfo{pages}{2273} (\bibinfo{year}{2022}),
  \eprint{2210.06220}.

\bibitem[{\citenamefont{{Sanati} et~al.}(2020)\citenamefont{{Sanati}, {Revaz},
  {Schober}, {Kunze}, and {Jablonka}}}]{2020A&A...643A..54S}
\bibinfo{author}{\bibfnamefont{M.}~\bibnamefont{{Sanati}}},
  \bibinfo{author}{\bibfnamefont{Y.}~\bibnamefont{{Revaz}}},
  \bibinfo{author}{\bibfnamefont{J.}~\bibnamefont{{Schober}}},
  \bibinfo{author}{\bibfnamefont{K.~E.} \bibnamefont{{Kunze}}},
  \bibnamefont{and}
  \bibinfo{author}{\bibfnamefont{P.}~\bibnamefont{{Jablonka}}},
  \bibinfo{journal}{A\&A} \textbf{\bibinfo{volume}{643}}, \bibinfo{eid}{A54}
  (\bibinfo{year}{2020}), \eprint{2005.05401}.

\end{thebibliography}
\end{document}